# Axiomatic Theories of Intermediate Phases (IP) and Ideal Stretched Exponential Relaxation (SER)

J. C. Phillips,

Dept. of Physics and Astronomy, Rutgers University, Piscataway, N. J., 08854-8019

**Abstract**

Minimalist theories of complex systems are broadly of two kinds: mean-field and axiomatic. So far all theories of properties absent from simple systems and intrinsic to complex systems, such as IP and SER, are axiomatic. SER is the prototypical complex temporal property of glasses, discovered by Kohlrausch 150 years ago, and now observed almost universally in microscopically homogeneous, complex non-equilibrium materials (strong network and fragile molecular glasses, polymers and copolymers, even electronic glasses). The Scher-Lax trap model (1973) is paradigmatic for electronic SER; for molecular SER Phillips (3RCS 1995) identified two "magic" shape fractions $\beta = 3/5$ and $3/7$, as confirmed by many later experiments here reviewed. In the dielectric SER frequency domain involving ion conduction, there are also special beta values for fused salts and glasses, slightly, but distinguishably, different because of the presence of a forcing electric field.

## 1. Introduction

One of the emerging fields of theory is realistic analysis of exponentially complex systems. Even with modern computers such systems in practice often present insuperable barriers to brute-force simulations. Worse still, in the few cases where the simulations do succeed in reproducing some features of experiment (such as SER), they may not identify the organizing principles that are responsible for physical properties. Thus one turns to axiomatic theories, where success is measured by the simplicity of the axioms and the breadth and accuracy of the results. Truly successful axiomatic theories contain few or no adjustable parameters.

SER is ubiquitous in glasses and deeply supercooled glass-forming liquids. It is sometimes naively described as evidence for a distribution of relaxation times, but this "explanation" begs the question of why this *particular* distribution is so widely observed, in the most favorable cases over as many as seven temporal decades. So far the only model that derives this *particular* distribution is the Scher-Lax trap model [1], originally studied in the 1970's by them and others (see the review [2]) as a model for dispersive electronic transport in amorphous semiconductors (such as a-Si:H). Their model has become standard in that field, and has been cited several thousand times.

In the Scher-Lax trap model the traps are idealized as fixed sinks, randomly distributed in configuration space. Entropic excitations diffuse freely to the sinks, where they disappear into "black holes" *without crossing an activation barrier*. Stretching occurs naturally, as the excitations nearest the sinks are the first depleted, and those furthest away must diffuse longer times to be captured. A by-product of the trap model is that it predicts a specific value for the shape fraction β of SER described by $\exp[-(t/\tau)^\beta]$, where τ is the usual relaxation time when β=1. This value is fixed by the dimensionalities of the diffusion equation; it is

$$\beta = d/(d+2) \tag{1}$$

where d is the effective dimensionality of the configuration space in which glassy relaxation takes place. So far it seems that the model's prediction is trivial because it defines one parameter in turns of another. Moreover, the model requires a homogenous material, a condition not easily satisfied in amorphous electronic materials. Indeed, in the 1970's there was also little relaxation data on homogeneous molecular glasses. However, by the 1990's, while the data on homogeneous electronic relaxation remained limited, a large, modern data base on *intrinsic* relaxation in *microscopically homogeneous* molecular glasses and supercooled liquids had emerged.

The modern molecular glass data base contains both striking and puzzling features [2]. The positive feature is that in many glasses $\beta = \beta_g = 3/5$, with an accuracy of a few %. From (1), this immediately gives d = 3, a gratifyingly simple result. The success of the model is even more striking in the light of the prediction of β = 1/3 by spin-glass models, the prototypical model for *on-lattice* relaxation [2]. The failures of brute-force lattice models to describe SER can often be traced to their inability to describe *off-lattice* configurational entanglement or knots, so one can reasonably infer that the traps postulated by the model correctly simulate the entropy sinks that are needed for relaxation. At the same time, these sinks or knots serve a second purpose, characteristic of the simplicity and economy crucial for axiomatic success: they can also be regarded as the key *off-lattice* structural obstacles that prevent crystallization. The failure of spin-glass models to give a satisfactory description of glassy relaxation lies in the way that these obstacles are artificially imposed through arbitrary interactions on an equally arbitrary lattice. The example of spin glasses shows the dangers to which axiomatic models are exposed when they are not well-grounded in physics; even when they do succeed qualitatively in obtaining SER, the results are still quantitatively unsatisfactory.

The great success of the Scher-Lax trap model in describing part of the modern molecular glass data base is apparently contradicted by its striking failures to explain the values of β for many other glasses, especially a-Se and many polymers. There one finds that the values of β are all clustered around β = 0.43. (Some well-known commercial polymers, like polyvinyl chloride (PVC) and polystyrene (PS) have much lower values of β, but it has been known since the 1930's that all these "worst cases" are partially crystallized and microscopically inhomogeneous (atactic) [2].) However, let us insert 0.43 in (1), and solve for d. The result is d = 3/2, and at this point one would have to be deaf not to hear what experiment is trying to tell us. While d = 3 suggests relaxation by density fluctuations, there are long-range uniaxial strain fields associated with the chain structures of polymers and a-Se. One can then suppose [2] that the glassy relaxation dynamics is affected both by short-range forces, which are effective, as well as long-range forces, which are ineffective, and define an effective or fractal dimensionality for competitive relaxation, which is

$$d^* = fd, \tag{2}$$

where f = ½ = (number of short-range forces)/(total number of forces) and

$$\beta = d^*/(d^* + 2) = d/(d + 2/f). \tag{3}$$

Equation (2) is a typical axiom: it is simple, but at the same time, like the configurational knots, it embodies some very deep and fundamental considerations. Note that although f is adjustable, it always has the form of a simple fraction, such as 1, 1/2 or perhaps 1/3.

Numerical studies have shown [2] that SER is an asymptotic property that sets in only after a transient simple exponential has died out. If one starts out with a mathematical model of sinks randomly embedded in an ideal (random) gas, then this asymptotic regime is reached only after astronomically long times [2], when the signal is exponentially small (~ $10^{-10}$ or less). However, simulations of good glass-formers typically reach the asymptotic regime when the signal is ~ 3/4 of the initial value [2]. The key to understanding this paradoxical result is that to avoid crystallization and reach a glass transition, one must *choose* a good glass-former to study, and such a good glass former retains only a small fraction (such as 1/4) of the truly "random" disorder contained in the mathematical gas model; the chosen model by selection (usually based on extensive numerical surveys) already contains a high level of configurational entanglement, designed to simulate a good glass-former.

In general one might have expected that the relative weights of effective and ineffective relaxation channels could be arbitrary, but this expectation, like the one for arbitrary distributions of relaxation times, is naïve and simplistic. If one channel were more lightly weighted than another, it would quickly "fill up", and the effective weights of the two would be equalized, until such time as perhaps a new relaxation channel, with a greatly different relaxation time, would appear. Over fairly wide time ranges the number of relaxation channels should be integral and they should be equally weighted; this is the principle of a priori equidistribution of kinetic microentropies [3].

## 2. Early Data Base

For historical convenience one divides the data base into two parts, the early part corresponding roughly to experimental papers published in the period 1986-1996, and papers published after 1996. The former are discussed in great detail in [2] and are only listed here, while the latter (which correspond to *predictions* of the axiomatic theory) are discussed here and in [3]. Values of β based on partially crystallized samples or otherwise obviously inhomogeneous samples (atactic polymers, for example, or home-made "composites"), or obtained from estimates based on shorter temporal ranges (typically two decades or less) are regarded as generally not reliable. Because of space limitations the reliable early data are not shown here, but they are collected in [2]: there one finds values of β for 9 inorganic network (strong) glasses, usually measured macroscopically by stress relaxation (quite slow, times longer than 1 s), Table 3 of [2]. Polymer relaxation is measured best using spin-polarized neutron scattering, a very sophisticated technique that produced a master curve exhibiting SER for *cis-trans* polybutadiene spanning more than seven decades in t/τ with β = 0.43(2), in perfect agreement with the value $\beta = 3/7$ predicted by Eqn. (3) with $f = ½$. The values of $\beta$ for 3 polymers obtained by spin-polarized neutron scattering agree well with those obtained by dielectric relaxation and are clustered near 0.43, while values for 5 other polymers were obtained by stress relaxation, and are usually close to 0.43, as shown in Table 4 of [2]. PIB and cis-PI are noteworthy exceptions, where bulky methyl sidegroups create density fluctuations; the upward shift of $\beta$ from 3/7 towards 3/5 scales with the methyl sidegroup density. Coulomb interactions usually facilitate crystallization, so the only fused salt where $\beta$ has been well-studied is KCN = $(KNO_3)_x (Ca(NO_3)_2)_{1-x}$, which is a good glass former for $1/3 < x < 2/3$. The results for different experiments are shown in Table 5 of [2], and the accompanying discussion shows why light probes, which couple to both long- and short-range forces, give $\beta$ close to 3/7, while neutron scattering, which couples only to short-range forces, gives $\beta$ close to 3/5.

Molecular mixtures generally phase separate and crystallize very rapidly; this does not occur in the popular molecular dynamics simulations of mixed spherical fluids only because the time spans of the simulations are so short (< 10 ns). Thus there are only a few good molecular glass formers; most of these are polar alcohols, with terminal OH ions providing steric hindrance to obstruct crystallization. The most-studied non-polar organic (fragile) glass is OTP (o-terphenyl), which is available commercially in high purity; as expected, its relaxation is dominated by density fluctuations and $\beta = 0.61(1)$. Values of $\beta$ for six organic fragile glasses are listed in Table 6 of [2], including nonpolar OTP and 5 polar alcohols. The 9 values, obtained by specific heat, ultrasonics, light scattering, and spin-polarized neutron scattering, are 0.61(1), showing that these probes couple to density fluctuations which give $\beta = 3/5$ (the same value as the strong network glasses!). Much larger values of $\beta$ are obtained in dielectric relaxation experiments and are listed for completeness in Table 6. It now appears that those values are unreliable because of the limited accessible frequency range and because of inadequate numerical analysis (the effects of double layers at the electrodes were ignored), as discussed correctly in [3] and below.

## 3. Recent Data: Many Successful Predictions!

SER continues to be probably the most demanding and instructive single dynamical phenomenon that can be observed in experiments on glasses: demanding, because of the requirement of microscopic sample homogeneity; instructive, because of the nearly always excellent separation of interaction channels into effective and ineffective. Recent data have shown that the predictions of the axiomatic Scher-Lax-Phillips (SLP) configurational sink model are remarkably successful [3]. There have been two studies of OTP, one by Brillouin scattering [4], which excites surface waves that relax by a mixture of short- and long-range forces, and these gave $\beta = 0.43$, as predicted. A new method, multi-dimensional NMR, enables one to study [5] the relaxation of states prepared over long and short times, so that relaxation can take place either with short-range forces only, or with a mixture of short-and long-range forces, leading to two values of $\beta$, 0.59 and 0.42, both in excellent agreement with the predicted values of 3/5 and 3/7. Photon correlation spectroscopy revealed [6] the differences between relaxation of poly(propylene glycol) (PPG) and PG in the wide time range $10^{-8}–10^{-4}$ s. In PPG hydroxyl terminals are separated by N organic monomers, and for $7 < N < 70$ these polymers exhibit a constant β nearly independent of N with β = 0.43, in good agreement with the value 0.42 obtained by stress measurements [2], and quite distinct from the PG value [2] measured by specific heat, β = 0.61. Thus the value of β predicted for mixed short-and long-range forces of 3/7 is confirmed for PPG, while PG itself exhibits relaxation by short-range forces only, with β = 3/5.

Pure polystyrene, which consists of the usual hydrocarbon main chain with a phenyl side group, is atactic (partially crystallized) and unsuited to study of intrinsic relaxation. However, commercially PS is stabilized, and presumably rendered microscopically homogeneous, by the addition of a suitable volume fraction of acrylonitrile. Analysis of the birefringence of stressed *commercial* poly(styrene-*co*-acrylonitrile) films [7] led to the identification of two relaxation processes, the fast one of the main chain, and the $10^3$ slower one of the phenyl side group. The main chain shows the same value of $β_I$ = 3/7 = 0.43 as Se and many polymer chains did in both experiments described above, while the phenyl side group gave $β_{II}$ = 0.32. The value of $β_I$ = 0.43 is just what we can expect from main chain relaxation with two channels and $f_I$ = 1/2, only one of which is relaxational, while the other is cyclical or ineffective. Obviously $β_{II}$ = 0.32 corresponds to $f_{II}$ = 1/3, that is, relaxation with three channels, only one of which is relaxational, while the other two are cyclical or ineffective. What is the origin of the extra ineffective channel for the slow relaxation of the phenyl side group? It is probably the interactions of the phenyl side group with the acrylonitrile copolymer. There are many commercial copolymers, and birefringence is commonly used industrially as a simple means of quality control. Axiomatic SLP theory shows that there is a firm scientific basis for this practice.

**4. Field-Forced Molecular (Ionic) Relaxation**

Traditionally values of β obtained from dielectric (ac frequency υ < 10 MHz) relaxation studies have been regarded as merely qualitative. However, in a long series of papers Macdonald carefully analyzed many sets of dielectric relaxation data, paying special attention to the elimination of corrections due to double layers at electrodes, etc., to obtain very accurate fits. Like all relaxation studies, his work has benefited from increased digitalization, and it eventually converged to a unique "best fitting" model with β = 1/3 in the low frequency limit. This value is uncertain to ± 0.01, and a similar uncertainty attaches to β = 3/5 (temporal relaxation by density fluctuations) and even to β = 3/7, so there is a clear-cut discrepancy here. This discrepancy is resolved by the recognition that in dielectric relaxation experiments relaxation does not occur, as in the temporal SLP model, by free diffusion of excitations to entropic sinks, but instead the excitations are entrained by the applied electric field, and relax relative to that field [3].

In a plane parallel capacitative geometry the field is normal to the planes, and the only effective coordinate is that parallel to the applied field, the two transverse coordinates being ineffective (f = 1/3). Thus d* = fd =1, and β = 1/3. In this case the temporal SLP model applies to field-forced relaxation, but β = 1/3 can also be obtained by arguing that this result follows from simple fractal arguments as well [3]. The value of 1/3 leads to a semi-universal frequency response model, the K1 [3], one that involves high-frequency power-law response of the real part of the conductivity with an exponent of 2/3, in agreement with many experimental results. It is important to note that 2/3 is also the high-frequency value obtained for a model directly involving SER with a beta of 2/3. Such a beta value follows immediately from the axiom that in the high-frequency limit relaxation occurs relative to the instantaneous local field, defined by z coordinate and polar angle, with only the azimuthal angle being ineffective [3]. This way of counting works equally well at low frequencies, where it gives $\beta = 1/3$, in agreement with the SLP field-free result. Thus the differences between field-forced and field-free relaxation are small (no difference in the low frequency limit) and at high frequencies, where only short-range forces might be important, the difference is only between 3/5 and 2/3. However, best fits clearly indicate 1/3 and 2/3 in the appropriate limits, so the field-forced case is observably different.

## 5. Electronic Relaxation

As remarked in the introduction, electronic relaxation in amorphous semiconductors generally occurs heterogeneously. In a-Si:H, at room temperature (the sample storage or relaxation temperature), measurements of H ion relaxation give a lower limit for $\beta \sim 0.43$ (3), which is reasonable, as relaxation in a semiconductor is likely to involve both traps and space charge (short- and long-range forces, respectively) [2]. Relaxation of photoinduced absorption in fullerene ($C_{60}$) similarly gave $\beta = 0.40$. Optical SER in Cd(S,Se) nanocrystallites embedded in a borosilicate matrix (used commercially as optical filters) have a rich variety of SER phenomena that are explained quite completely in terms of variable effective dimensionalities [2], with the smallest value of $\beta \sim 0.4$. Qualitatively similar SER occurs in luminescent porous Si with a range of $\beta$ between 1 and 0.4 [8] (another predictive success!) where a dynamical metal-insulator transition is marked by a rapid increase in $\beta$ above 2.6 eV. It appears that in each material in some spectral ranges the excited carriers or ions form a metastable Coulomb glass with $f = 1/2$.

Near a metal-insulator transition in an irregularly doped crystal, however, intrinsic results characteristic of homogeneity might be obtainable. Thus SER occurs in the context of photoinduced (super)conductivity in the cuprates (specifically the very well-studied $YBa_2Cu_3O_x$ at the metal-insulator transition, x = 6.4, where ideal (microscopically homogeneous) percolative electronic networks are formed) [3]. The superconductive cuprates are layered pseudoperovskites, with in-plane conductivities 10-100 times larger than perpendicular conductivities; is $\beta$ also anisotropic? It certainly doesn't do any harm to look at the experimental values, which cluster around $\beta$ (in plane) = 0.56(1) and $\beta$ (out of plane) = 0.59 (1). The differences are small, but possibly significant, and are explained by axiomatic theory as reflecting 3/5 (field screened) and 2/3 (field-forced) SER, respectively [3]. In any event, it appears that in the cuprates the electronic values of $\beta$ are reasonable, and in generally good agreement with the SLP predictions.

## 6. Extrinsic Relaxation

The SLP no-parameter theory of SER stresses diffusive intrinsic relaxation to randomly distributed fixed sinks in a microscopically homogeneous medium. There are relatively few samples that are close enough to ideal to be described as microscopically homogeneous. What can be said of the many other extrinsic cases? First, one can note that if experiments are to be done using sophisticated probes (for example, spin-polarized neutrons), then no effort should be spared to obtain samples of the highest possible quality; in fact, this rule has always been observed for those experiments. Second, even if one is doing a relatively simple experiment, such as stress relaxation of films measured by birefringence, one can still hope to obtain spectacular intrinsic results by judicious sample selection. Suppose, however, that one is interested mainly in a "quick and dirty" relaxation study of an extrinsic sample that happens to be at hand. Some kind of relaxation is observed, and it is not simple exponential. In this case often an adequate fit is obtained with two exponentials, which is credible if the two relaxation times differ by more than an order of magnitude. If not, one may want to use a qualitative model to describe residual non-exponential relaxation.

Popular models [9,10] of extrinsic inhomogeneous relaxation generally begin with the Gibbs (1880) droplet nucleation model of first order phase transitions, which balances a negative volume free energy difference against a positive interfacial energy. Relaxation occurs *via* Arrhenius (1884) activation over independent nucleation barriers. To simulate "randomness" these energies are initially assumed to obey a displaced Gaussian distribution. It was shown some time ago [11] that the predictions of a simple exponential distribution of relaxation times or activation energies could fit those of a Gaussian distribution quite well over most of the experimentally accessible range, but neither distribution is a good approximation for the actual distribution associated with SER, one that is available in accurate numerical form for any appropriate beta value in the free LEVM computer program [3].

An approximate exponential distribution of (fictive) activation energies for SER, derived by steepest descent and then further approximated, is [12]

$$P(E) \sim \exp\{C(\beta)\exp[-\alpha(E - E_0)]\} \qquad (4)$$

with $\alpha = \beta/(1 - \beta)$. This monotonically decreasing function is never similar to a displaced Gaussian $\exp[-\alpha(E - E_0)^2]\}$, or even to an exponentially iterated displaced Gaussian. Attempts to derive SER from Gaussian-fitted activation energies produce large errors (for instance, the fitted value of β derived from a series expansion of a Gaussian fit to a wide distribution of activation energies obtained from a simulation of SER with $\beta = 0.61$ is only $\beta = \beta_G = 0.41$ [10]). Perhaps one can correct for this failing to some (unknown) extent by collapsing *ad hoc* the upper half of the Gaussian distribution to its initial mean value [9]. The resulting theory [9] has the interesting feature that it correlates some specific heat jumps with β well for partially crystallized materials (PVC, PS, $GeO_2$), while it fits badly true glasses (like $SiO_2$). It appears that through multiply uncontrolled and largely untested approximations, qualitative descriptions of extrinsic relaxation are obtainable, although the meaning of the postulated activation energies (and their distribution) is cloudy. Such models may be useful in biophysical applications where the data are meager and sample homogeneity is not achievable, although in those cases the extrinsic models will often contain many adjustable parameters.

## 7. Conclusions

The SLP model [2] has achieved many predictive successes for field-free temporal relaxation, and its extension to field-forced dielectric relaxation has consolidated many years of careful analysis of digitalized data [3]. The advantages of the abstract axiomatic approach over more concrete mean-field models are increasing. These advantages should appeal strongly to theorists interested in analyzing experimental data [13].